\documentclass[useAMS,usenatbib]{mn2e}

\usepackage{mathptmx}
\usepackage{multirow}
\usepackage{psfrag}
\usepackage{pspicture}
\usepackage{graphicx}
\usepackage{amsmath}

\title[The highest-frequency detection of a radio relic]{The highest-frequency detection of a radio relic: 16-GHz AMI observations of the `Sausage' cluster}
\author[A. Stroe et al.]{Andra Stroe$^{1}$\thanks{E-mail: astroe@strw.leidenuniv.nl}, Clare Rumsey$^{2}$, Jeremy J. Harwood$^{3}$, Reinout van Weeren$^{4}$,
\newauthor Huub J. A. R\"ottgering$^{1}$, Richard D. E. Saunders$^{2,5}$, David Sobral$^{1}$,
Yvette C. Perrott$^{2}$, 
\newauthor Michel P. Schammel$^{2,6}$ \\
$^{1}$Leiden Observatory, Leiden University, P.O.\ Box 9513, NL-2300 RA Leiden, The Netherlands\\
$^{2}$Astrophysics Group, Cavendish Laboratory, JJ Thomson Avenue, Cambridge, CB3 0HE\\
$^{3}$School of Physics, Astronomy and Mathematics, University of Hertfordshire, College Lane, Hatfield, Hertfordshire AL10 9AB, UK\\
$^{4}$Harvard Smithsonian Center for Astrophysics (CfA - SAO), 60 Garden Street Cambridge, MA 02138, US\\
$^{5}$Kavli Institute for Cosmology Cambridge, Madingley Road, Cambridge CB3 0HA, UK\\
$^{6}$I.N.A.F. - Osservatorio Astronomico di Roma, via Frascati 33, 00040 - Monte Porzio Catone (Roma), Italy\\
\vspace{-15pt}
}
\begin{document}
\maketitle
\begin{abstract}
We observed the cluster CIZA J2242.8+5301 with the Arcminute Microkelvin Imager at $16$ GHz and present the first high radio-frequency detection of diffuse, non-thermal cluster emission. This cluster hosts a variety of bright, extended, steep-spectrum synchrotron-emitting radio sources, associated with the intra-cluster medium, called radio relics. Most notably, the northern, Mpc-wide, narrow relic provides strong evidence for diffusive shock acceleration in clusters. We detect a puzzling, flat-spectrum, diffuse extension of the southern relic, which is not visible in the lower radio-frequency maps. The northern radio relic is unequivocally detected and measures an integrated flux of $1.2\pm0.3$ mJy. While the low-frequency ($<2$ GHz) spectrum of the northern relic is well represented by a power-law, it clearly steepens towards $16$ GHz. This result is inconsistent with diffusive shock acceleration predictions of ageing plasma behind a uniform shock front. The steepening could be caused by an inhomogeneous medium with temperature/density gradients or by lower acceleration efficiencies of high energy electrons. Further modelling is necessary to explain the observed spectrum.
\end{abstract}
\begin{keywords}
acceleration of particles, radiation mechanisms: non-thermal, shock waves, galaxies: clusters: individual: CIZA J2242.8+5301, radio continuum: general
\vspace{-15pt}
\end{keywords}
\section{Introduction}\label{sec:intro}
Radio relics are diffuse, strongly-polarised, Mpc-wide synchrotron objects found at the periphery of disturbed galaxy clusters \citep[e.g.][]{2001A&A...373..106F}. Relics are thought to trace large-scale, fast, outward-travelling shock fronts (Mach numbers up to $4$) induced by major mergers between massive clusters \citep{1998A&A...332..395E, 2002ASSL..272....1S, 2012A&ARv..20...54F}. These objects usually extend perpendicularly to the merger axis of their host cluster and display narrow transverse sizes, resulting from a spherical-cap-shaped regions of diffuse emission seen side-on in projection \citep{2012A&ARv..20...54F}. Integrated radio spectral indices of elongated relics below $<1.2$ GHz range between $-1.6<\alpha<-1.0$ ($F_{\nu}\propto\nu^{\alpha}$) and the spectra display no curvature up to $\sim2$ GHz \citep{2012A&ARv..20...54F}. \citet{1998A&A...332..395E} suggest relics are formed through the diffusive shock acceleration mechanism \citep[DSA; e.g., ][]{1983RPPh...46..973D}. In this scenario, intra-cluster-medium (ICM) particles are accelerated by shocks to relativistic speeds in the presence of $\mu$G level magnetic fields at the outskirts of clusters \citep[e.g.][]{2009A&A...503..707B, 2010A&A...513A..30B}. Due to low acceleration efficiencies, mildly-relativistic (rather than thermal) electrons likely cross the shock surface multiple times by diffusing back through the shock after each passage. These re-accelerated electrons then exhibit  synchrotron radio emission. 

CIZA J2242.8+5301 \citep[`Sausage' cluster;][]{2007ApJ...662..224K, 2010Sci...330..347V} hosts a remarkable example of double, Mpc-wide, narrow radio relics. Twin relics are thought to form after a head-on collision of two roughly equal-mass clusters \citep{1999ApJ...518..603R}. The northern relic (RN) is bright ($0.15$ Jy at $1.4$ GHz) with an integrated spectral index between $153$ MHz and $2.3$ GHz of $\alpha_\mathrm{int}=1.06\pm0.04$ \citep{2013A&A...555A.110S}. RN displays spectral index steepening and increasing curvature from the outer edge of the relic towards the inner edge, thought to be due to synchrotron and inverse Compton losses in the downstream area of a shock with an injection spectral index of $\sim-0.65$. The cluster contains a fainter counter-relic towards the south, a variety of diffuse patches of emission and a number of radio head-tail galaxies \citep{2013A&A...555A.110S}. 

Relics have been primarily studied at low radio frequencies ($<1.5$ GHz), making accurate determination of the injection, acceleration and loss mechanisms difficult. Most of the $\sim40$ radio relics with published spectra \citep{2012A&ARv..20...54F} have measurements up to $2.3$ GHz, while only two relics have spectra derived up to $5$ GHz \citep[Abell 521, 2163;][]{2008A&A...486..347G, 2001A&A...373..106F}. The scarcity of high radio-frequency observations of relics is caused by two factors: (i) the steep spectrum means that relics are significantly fainter at high frequencies; (ii) there are few radio telescopes with the required compact uv coverage needed to detect relics. To begin to address this, we performed exploratory observations at $16$ GHz with the Arcminute Microkelvin Imager \citep[AMI;][]{2008MNRAS.391.1545Z} of the `Sausage' cluster. AMI is the only cm-wavelength radio telescope with the required capabilities for detecting Mpc-wide, low-redshift, diffuse targets at sub-arcminute resolution. In this letter, using two different AMI configurations, we image the `Sausage' cluster at high ($40$ arcsec) and low ($3$ arcmin) resolutions. By combining the data with measurements from the Giant Metrewave Radio Telescope (GMRT) and the Westerbork Synthesis Radio Telescope (WSRT), we derive the RN spectrum over the widest frequency coverage ever performed for a radio relic (between $153$ MHz and $16$ GHz) and compare our results with predictions from spectral-ageing models. At the redshift of the `Sausage' cluster, $z=0.192$, $1$ arcmin corresponds to a scale of $0.191$~Mpc. All images are in the J2000 coordinate system. 
\vspace{-10pt}
\section{Observations \& Data Reduction}
\label{sec:obs-reduction}
For our analysis, we combine the existing WSRT and GMRT observations with new AMI observations. We use the WSRT and GMRT datasets presented in \citet{2013A&A...555A.110S} and refer the reader to that paper for details of the data reduction. In summary, the data were flagged, bandpass and gain calibrated and bright sources in the field were removed using the `peeling' technique \citep{2004SPIE.5489..817N}. A total of three frequencies were observed with the GMRT: $153$, $323$ and $608$ MHz and four with the WSRT: $1.2$, $1.4$, $1.7$ and $2.3$ GHz. 
\vspace{-10pt}
\subsection{AMI observations}
\label{sec:obs-reduction:AMI}
AMI is a dual array of interferometers located near Cambridge, UK. The Small Array (SA) and the Large Array (LA), observe over $13.9-18.2$ GHz and measure the single Stokes polarisation I+Q \citep{2008MNRAS.391.1545Z}. The SA has ten $3.7$-m antennas with baselines of $5-20$ m, while the LA has eight $12.8$-m antennas with baselines of $18-110$ m, giving the instruments sensitivities to complimentary ranges of angular scale. Observations towards the X-ray cluster centre were taken between July 2012 and February 2013 on both SA and LA, with the field observed with a single pointing with the SA and with a series of multi-point hexagonal raster observations on the LA. The northern relic itself was also observed with the LA with four pointings centred along its axis. For all observations, flux calibration was performed using observations of 3C\;48, 3C\;286 and 3C\;147, with 3C\;286 calibrated against VLA measurements \citep{2013ApJS..204...19P}. Raw data were flagged for hardware errors, shadowing and interference and phase and amplitude calibrated using the in-house software package {\sc reduce} \citep{2009MNRAS.400..984D}. All of the reduced LA data were concatenated into a single uv dataset before mapping.

Unlike the WSRT and GMRT arrays, that measure total intensity I, both AMI arrays measure the single polarisation Stokes I+Q. For the spectral work, it is necessary to correct the AMI values to make them consistent with those from the other telescopes. $4.9$-GHz observations \citep{2010Sci...330..347V} show that RN is $60\%$ polarised and that the magnetic field is tangent to the relic. The measured Faraday depth of $-140$ rad m$^{-2}$ implies a rotation of $24$ degrees between $5$ and $16$ GHz. We assume the same degree of polarisation at $16$ GHz as at $4.9$ GHz. AMI calibration assumes random polarisation, which in the case of RN is correct for only $40\%$ of the measured flux. We multiply $60\%$ of the measured I+Q flux by $1/(2 \cos^2\phi)$, where $\phi$ is the angle between the electric field vector and the orientation of the I+Q AMI feeds, which is vertical on the sky. We take into account the variation of $\phi$ along the relic. The unpolarised RN flux is added to the corrected polarised flux: $I=40\%(I+Q)+60\%(I+Q)/(2 \cos^2\phi)$. The I integrated flux density is obtained by decreasing $24\%$ from the I+Q value.
\vspace{-10pt}
\subsection{Imaging}
\subsubsection{AMI radio maps}
\begin{figure*}
\begin{center}
\includegraphics[trim=0cm 0cm 0cm 0cm, width=0.495\textwidth]{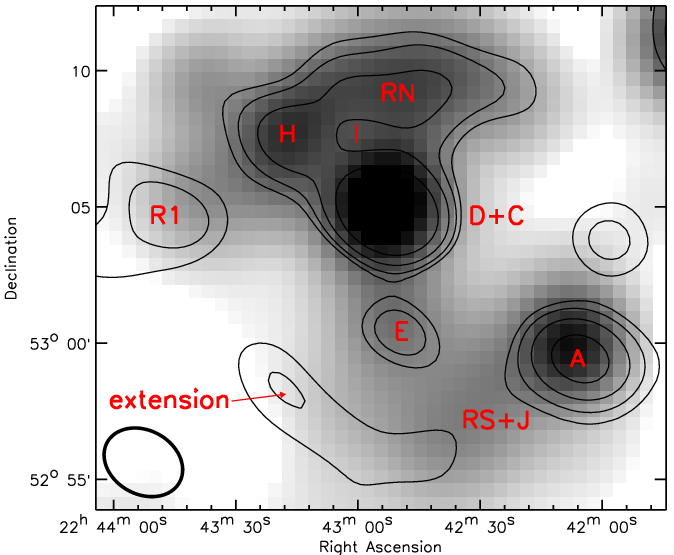}
\includegraphics[trim=0cm 0cm 0cm 0cm, width=0.495\textwidth]{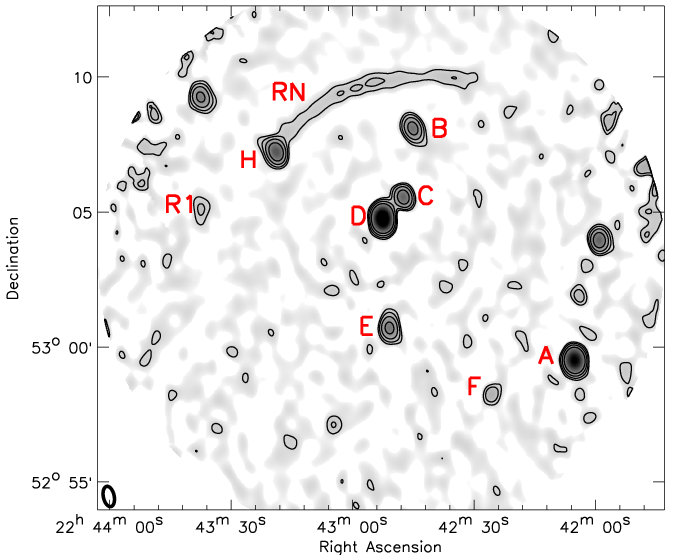}
\end{center}
\vspace{-10pt}
\caption{`Briggs'-weighted AMI $16$~GHz images (robust=$2$). Contours drawn at ${[4,8,16,32]} \times \sigma_{\mathrm{RMS}}$. \textit{Left}: AMI SA. The beam size $3.0$ arcmin $\times2.3$ arcmin is shown in the bottom-left corner of the image. The noise is $\sim0.1$ mJy beam$^{-1}$. The grey intensity shows a low-resolution ($3$ arcmin) WSRT $1.5$ GHz image. \textit{Right}: AMI LA in intensity and contours, at $44$ arcsec $\times22$ arcsec resolution, with $\sigma_\mathrm{RMS}\approx35$ $\mu$Jy beam$^{-1}$. Source labelling from \citet{2013A&A...555A.110S} is shown.}
\label{fig:map}
\vspace{-10pt}
\end{figure*}

Figure \ref{fig:map} shows separate CLEANed maps for the LA and SA data, using `Briggs' weighting \citep[Robust set to 2.0 to enhance diffuse emission,][]{briggs_phd}. The SA map resolution is $3.0$~arcmin $\times 2.3$ arcmin, while the LA has $44$ arcsec $\times 22$ arcsec. The RMS noise in the SA map is $\sim0.1$ mJy beam$^{-1}$ near the northern radio relic, while in the LA map it is $35\mu$Jy beam$^{-1}$.
\vspace{-10pt}
\subsubsection{Combining the GMRT, WSRT and the AMI radio maps}
To produce directly-comparable, multi-frequency radio images, a number of steps were taken before combining the maps for the study of the integrated spectrum. Due to the very low SA resolution compared to the WSRT and GMRT maps, we chose to combine only the LA map with the other datasets. We imaged the data using the CLEAN algorithm with the same pixel size ($1$ arcsec per pixel), image size and uniform weighting. The uv-coverage of the LA samples densely down to a uv-distance of $0.8$ k$\lambda$. Therefore, only GMRT and WSRT data beyond a uv distance of $0.8$ k$\lambda$ were used, so that our radio maps image approximately the same spatial scales on the sky. We simulated an LA observation of a uniform brightness distribution with the angular dimensions of the northern relic as measured by the LA, with the uv coverage and pointings used for the real LA observation. We found that the LA could be resolving out a negligible part of the largest scale diffuse emission. The uv-cut is necessary for extended sources, as inconsistent inner-uv coverages can lead to non-comparable integrated fluxes. All of the maps were primary beam corrected and convolved to the beam of the AMI LA map.

We adopt an absolute flux-scale uncertainty of $10$ per cent for the GMRT and WSRT data, following \citet{2013A&A...555A.110S}. This uncertainty results from telescope pointing errors and imperfect calibration. AMI flux scale errors are well-described by $5$ per cent of the flux \citep{2011MNRAS.415.2708A}.
\vspace{-10pt}
\section{Results}
\label{sec:results}
\subsection{Radio morphologies}
\subsubsection{SA map}
The left panel of Figure~\ref{fig:map} shows a $1.5$-GHz WSRT image convolved to the resolution of the SA map, overlaid with SA contours. AMI recovers most of the bright sources detected by WSRT, while the fainter sources are below the noise. Due to the low resolution, sources close to the SA map centre will not only be blended and merged together, but will also have a negative contribution from SZ signal of the cluster. However, because of the excellent uv-coverage at short baselines down to $0.2$ k$\lambda$, all of the diffuse emission visible at lower frequencies is recovered in the AMI SA map.

The northern relic (RN) displays an arc shape, but the emission is mixed with radio galaxies H and B and diffuse source I (see also right panel, Figure~\ref{fig:map}).

The complex of diffuse emission towards the south of the cluster arises as a blending of sources RS, J, A and tailed-radio source F. At lower frequencies, radio phoenix J is much brighter than the relic RS \citep{2013A&A...555A.110S}. Since J has a much steeper spectrum than RS, they contribute comparably to the flux at $16$ GHz. Puzzling is the $\sim2$ mJy integrated flux, $1$ Mpc, diffuse extension of RS towards the east, which has no counterpart in the low-resolution WSRT map (see labelling in Figure \ref{fig:map}). By placing a $3\sigma$ upper limit on the WSRT flux (giving $\sim15$ mJy), we would expect the spectral index of this extension to be flatter than $-0.5$. While the peak in the WSRT emission is towards the west, at the location of compact radio galaxy A, and it progressively wanes towards the east, the AMI SA emission shows the opposite trend. The peak at the emission is located where no counterpart is seen in the WSRT map. There could be some point-source contamination, but this should be minimal as the LA finds no significant sources in the area.
\vspace{-10pt}
\subsubsection{LA map}
The right panel of  Figure \ref{fig:map} shows the AMI LA map (imaged with robust=2). The higher resolution enables a better deblending of sources, but the poorer inner uv-coverage leads to loss of flux on large scales. This is evident as most diffuse sources (RS, J) detected in the SA effectively disappear in the LA map. 

The northern relic is detected at the $11\sigma$ level at peak flux and clearly separated from its neighbouring source H towards the west. Only the central, brightest part of source R1 is visible. We also detect sources labelled A, B, C, D and E as point sources with high S/N ($>32\sigma$). The nucleus of tailed-radio galaxy F is detected at $10\sigma$, but its steep spectrum tail is not recovered, as expected \citep[see][]{2013A&A...555A.110S}. The `extension' is not detected in the high-resolution $16$ GHz, suggesting it may have a diffuse nature. 
\vspace{-10pt}
\subsection{Integrated spectrum}\label{sec:results:integrated}
Figure \ref{fig:intspec} and Table~\ref{tab:int_flux} present the spectrum of RN. The flux densities are measured in fixed boxes in uniform-weighted maps. Note that because of the uniform weighting, RN is detected at $6\sigma$ level significance at peak. We use a least-squares method to fit a single power law to the integrated flux-density of the relic from each of the eight radio maps, at common resolution and with the common uv-cut. This fitting takes into account a total flux error computed as the quadrature of the flux scale error of $10$ per cent for the GMRT and WSRT measurements and $5$ per cent in the AMI LA, and the RMS noise in each map multiplied by the square root of the number of beams contained in the box we measure the flux in. 

From spatially-resolved, low-frequency observations of RN, we found a $\sim-0.6$ injection index, with an integrated spectrum between $153$ MHz and $2.3$ GHz well-described by a linear fit with slope $-1.06$ \citep{2013A&A...555A.110S}. Figure~\ref{fig:intspec} shows in the dotted line the injection spectrum of the freshly-accelerated electrons, while the dashed line presents the integrated spectrum, as derived from the low-frequency data. A single power law fit ($\alpha_\mathrm{int}=-1.33\pm0.03$) provides a poor description of the data up to $16$ GHz, as the fitted line fails to pass through all but two error bars, with a reduced $\chi^2_\mathrm{red}$ of $163$ (solid line in Figure~\ref{fig:intspec}). The $16$ GHz measurement lies $12\sigma$ below the extrapolation of the low-frequency spectrum.
\vspace{-10pt}
\section{Discussion}\label{sec:discussion}
Radio relics are thought to form at the wakes of travelling shock fronts produced by the major merger of galaxy clusters \citep{2012A&ARv..20...54F}. The physical processes underlying their formation, such as the injection and ageing mechanism, can be constrained using high-frequency measurements, which have not been performed until now. Here, we present the $16$ GHz measurement of a relic through AMI observations of the `Sausage' cluster.
\vspace{-10pt}
\subsection{Northern relic}\label{sec:discussion:RN}
\begin{figure}  
\begin{center}
\includegraphics[trim=0cm 0cm 0cm 0cm, width=0.485\textwidth]{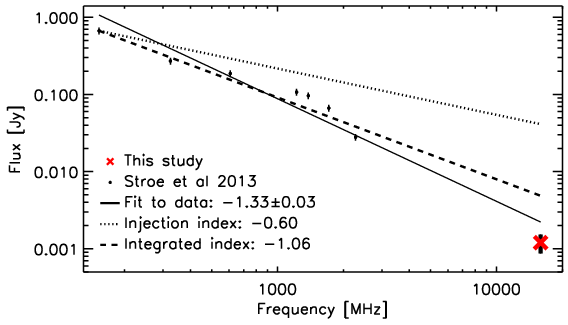}
\end{center}
\vspace{-10pt}
\caption{Integrated radio spectrum of the northern relic from $153$~MHz up to $16$~GHz (see also Table \ref{tab:int_flux}). The red cross marks the $16$ GHz measurement. The uncertainties include a $10$ per cent flux scale error added in quadrature to the $\sigma_\mathrm{RMS}$. A power-law is fitted to the eight frequencies. The dotted line shows the injection spectrum and the dashed line the integrated spectrum, as derived from high-resolution GMRT and WSRT data \citep{2013A&A...555A.110S}. The injection and integrated spectra below $2.3$ GHz are consistent with a CI model. The $16$ GHz measurement is $12\sigma$ below the CI prediction.}
\vspace{-10pt}
\label{fig:intspec}
\end{figure}

The $16$ GHz measurement of the northern relic and its integrated spectrum are given in Fig.~\ref{fig:intspec} and Table~\ref{tab:int_flux}. We find strong evidence for high-frequency steepening in the integrated spectrum of RN. There are two reasons why this should be considered a robust measurement. Firstly, the integrated spectra of point sources in the GMRT, WSRT and AMI LA maps are well described by single power laws, implying a correct overall flux scale also for the $16$ GHz measurements. Secondly, the dense AMI LA uv-coverage at the shortest spacings indicates minimal loss of flux at large spacial scales. 
\begin{table}
\begin{center}
\caption{Integrated radio spectrum of the RN measured in the uniform-weighted radio maps with common uv-cut and resolution. The uncertainties of the measurements are computed as the quadrature of the flux error and the rms noise in each map, multiplied by the square root of the number of beams spanned by the source. Note that RN is detected at a total S/N of $24$ in the integrated spectrum. Taking the lower bound given by the error in the integrated flux results in a $18\sigma$ detection.}
\vspace{-10pt}
\begin{tabular}{l c c c  c c c c c}
\hline
\hline
Freq. [GHz] & $0.15$ & $0.32$ & $0.6$ & $1.2$ & $1.4$ & $1.7$ & $2.3$ & $16$ \\
\hline
Flux [mJy] & $668$ & $270$  & $187$ & $107$  & $96$ &  $67$ & $28$ & $1.2$ \\
Error [mJy] & $69$ & $28$ & $19$ & $11$ & $10$ & $7$ & $3$ & $0.3$ \\ 
\hline
\end{tabular}
\label{tab:int_flux}
\vspace{-10pt}
\end{center}
\end{table}
All of the lower-frequency measurements (GMRT and WSRT, $<2.3$ GHz) present firm evidence for a scenario where the source traces an outward travelling shock wave. The ICM electrons are accelerated at the shock via the DSA mechanism, resulting in a relatively flat injection spectral index \citep[$\alpha\sim-0.6$;][]{2010Sci...330..347V, 2013A&A...555A.110S}. Energy losses due to synchrotron and inverse Compton processes lead to spectral index steepening and increasing spectral curvature in the downstream area \citep{2010Sci...330..347V, 2013A&A...555A.110S}. 

\citet{1998A&A...332..395E} modelled the integrated radio spectrum for such a relic formation scenario. At the shock front the particles are accelerated to a power-law radio spectrum, followed by losses that steepen the spectra. The integrated spectrum results from the summation of particle spectra spanning a range of ages from different regions in the downstream area. This is equivalent to the continuous injection model which was proposed to explain the integrated spectra of radio galaxies, where the jet deposits freshly accelerated electron in the radio lobes at a constant rate \citep[CI;][]{pacholcyzk}. 

In the CI model, the integrated spectrum has a critical frequency $\nu_\mathrm{crit}$, beyond which the spectrum steepens by $0.5$ because of energetic losses \citep{pacholcyzk}. \citet{1998A&A...332..395E} follows this approach and assumes that the integrated spectrum is measured beyond this critical frequency ($\sim100$ MHz), where we observe the aged spectrum. Therefore, simple plane-shock theory in the context of DSA predicts that the relic integrated spectral index of a source should be $0.5$ steeper than the injection index of the freshly-accelerated electrons, which has a hard upper limit at $-0.5$ \citep{pacholcyzk}. In \citet{2013A&A...555A.110S}, we showed that the spectral index and curvature maps for the northern relic are consistent with this model. The relic injection index is $-0.6$, which defines a shock front Mach number of $4.6\pm1.1$ \citep{2013A&A...555A.110S}. The difference between the injection index $\sim-0.6$ and the integrated index below $2.3$ GHz of $-1.06\pm0.04$ is consistent with the prediction from the CI model \citep{2013A&A...555A.110S}.

However, by extrapolating the RN low-frequency spectrum, we find that the $16$ GHz measurement is in stringent tension with the CI prediction, at the $12\sigma$ significance level (see Fig.~\ref{fig:intspec}). The integrated $153$ MHz to $16$ GHz index is much steeper ($\sim0.8$) than the injection index, while if only the high-frequency data is considered, this difference increases to $1.2$ spectral index units.
There are several explanations for this discrepancy between the $16$ GHz and the lower frequency measurements:
\begin{itemize}
\item As mentioned previously, the \citet{1998A&A...332..395E} model only holds for frequencies above the break frequency, where there is a balance between continuously, freshly injected plasma and ageing. In reality, there is a broad frequency range over which the steepening takes place. If the break occurs over a range of frequencies below $\sim100$ MHz, then the steepening would gradually increase towards higher frequencies, giving a curved integrated spectrum, as the one we observe in the northern relic. In cases where the spectral break occurs across the observed frequency range, the results will be biased to flatter integrated spectra and hence stronger derived Mach number. 
\item The injection spectrum is not a power law. With a pool of thermal or pre-accelerated electrons, the injection spectrum is still expected to be a power law \citep{2014arXiv1401.7519B}. The acceleration efficiency for electrons beyond $\gamma\approx3\times 10^4$ Lorentz factors (equivalent to a few GHz-peak emission frequency for $\mu$G magnetic fields) might be smaller than for the lower energy electrons. During their multiple crossings of the shock front, the electrons lose energy and radiatively cool during the acceleration, leading to a curved injection spectrum, assuming that the electron mean free path is larger than its gyro-radius \citep{2003ApJ...585..128K}.
\item A gradient of density and/or temperature across the source would result in different spectral components, resulting in a $16$-GHz spectrum completely dominated by losses/aged electrons. Assuming an isothermal sphere ICM gas distribution \citep{2002ASSL..272....1S}, we calculate a drop of $15-25$ per cent in electron gas density across the $50-100$-kpc width of the northern relic. The outward-movement of the shock, from regions of higher densities into lower densities, suggests that, in the past, the shock was crossing a region of higher electron density. The shock might have injected a larger pool of electrons in the past, compared to now. When summing up the particle spectra, the older electrons would have a higher normalisation, hence larger weight in the integrated spectrum. Therefore, the integrated spectrum would be dominated by the heavily-curved spectrum of the aged electrons. 
\item The magnetic field at the shock location might be stronger than in the downstream area, as a result of shock compression. Acceleration in the presence of ordered, strong magnetic fields at the shock front, combined with turbulent, lower magnetic fields in the downstream area, could lead to a curved integrated spectrum. Simulations of supernova remnant synchrotron emission under turbulent magnetic field conditions suggest that electrons in the cut-off regime can radiate efficiently \citep{2008ApJ...689L.133B}.
\end{itemize}
Nevertheless, higher-resolution data is required for distinguishing between these scenarios. At the moment, no relic formation mechanism can readily explain the high-frequency steepening, thus new theoretical models have to be developed \citep{2014arXiv1401.7519B}.
\vspace{-10pt}
\subsection{Diffuse extension}\label{sec:discussion:extension}
Towards the south of the cluster, we discover an extension at the $8\sigma$ significance level towards the east of RS in the low-resolution SA AMI map. This source does not have a counterpart in the lower frequency data, or in the high-resolution AMI LA map (Figure \ref{fig:map}), excluding the possibility of a point source. The extension appears elongated ($\sim1$ Mpc) and has a spectrum flatter than $\sim-0.5$. Its arc-like shape and proximity to RS make the extension an ideal candidate for a relic, but its flat integrated spectral index means that the source cannot result from a shock front in the context of DSA. Striking also is the difference between the spectral index of the extension and RS, which points to very different shock properties towards the south and towards the south-west. This could be explained by different ICM temperature/densities in the two directions. \citet{2013MNRAS.429.2617O} measured a sharp increase in ICM temperature in the direction of this extension, followed by a putative shock with a Mach number of $~1.2$, coincident with the location of the radio extension. Such an increase in temperature in the downstream area of travelling shock fronts has been also found in simulations \citep{rottiger1997}. The source seems to trace an arc-like shock front, which suggests a shock seen in projection onto the plane of the sky, which means the radio emission detected is a mixture of different age-populations of electrons. 
\vspace{-10pt}
\section{Conclusions}
High radio-frequency observations of steep-spectrum, diffuse, cluster emission have not previously been made owing to a lack of suitable instrumentation. We have observed the `Sausage' merging cluster at $16$ GHz at low ($3$ arcmin) and high ($40$ arcsec) resolution with the AMI array and we successfully detect diffuse radio relic emission for the first time at frequencies beyond $5$ GHz. Our main results are:
\begin{itemize}
\item The northern relic measures an integrated flux density of $1.2\pm0.3$ mJy ($6\sigma$ peak detection in a uniformly-weighted map). We investigate in detail its integrated spectrum and conclude there are clear signs of spectral steepening at high frequencies. If thermal electrons are accelerated, the steepening can be caused by a lower acceleration efficiency for the high-energy ($\gamma>3\times10^4$) electrons, a negative ICM density/temperature gradient across the source or turbulent downstream magnetic fields amplifying the emission of electrons in the cut-off regime. However, these scenarios are unlikely because of low-acceleration efficiencies at weak-Mach-number shocks. Further theoretical modelling is required.
\item We also detect a peculiar, flat-spectrum ($\alpha_\mathrm{int}\approx-0.5$) patch of diffuse emission towards the south-east of the cluster, which cannot be explained by the CI model.
\end{itemize}
The surprising high-frequency spectral steepening results and flat-spectra presented here suggest that the simple CI model, which has been widely used in the literature to explain the formation of radio relics, needs to be revisited. Furthermore, there is a clear need for high-quality radio observations of relics at cm and mm-wavelengths that resolve radio relics. 
\vspace{-12pt}
\section*{Acknowledgments}
We thank the referee for the comments which greatly improved the clarity and interpretation of the results. We also thank Gianfranco Brunetti, Tom Jones, Martin Hardcastle, Andrei Bykov, Matthias Hoeft, Wendy Williams and Marja Seidel for useful discussions. We thank the staff of the Mullard Radio Astronomy Observatory for their invaluable assistance in the operation of AMI, which is supported by Cambridge University. This research has made use of the NASA/IPAC Extragalactic Database (NED) which is operated by the Jet Propulsion Laboratory, California Institute of Technology, under contract with the National Aeronautics and Space Administration. This research has made use of NASA's Astrophysics Data System. AS acknowledges financial support from NWO. CR acknowledges the support of STFC studentships. JJH thanks the University of Hertfordshire and the STFC for their funding. RJvW is supported by NASA through the Einstein Postdoctoral grant number PF2-130104 awarded by the Chandra X-ray Center, which is operated by the Smithsonian Astrophysical Observatory for NASA under contract NAS8-03060. DS is supported by a VENI fellowship. YCP acknowledges the support of a Rutherford Foundation/CCT/Cavendish Laboratory studentship.
\vspace{-12pt}
\bibliographystyle{mn2e.bst}
\bibliography{AMI_sausage_final}

\begin{thebibliography}{}

\bibitem[AMI Consortium: Davies et al.(2009)]{2009MNRAS.400..984D}
Davies, M.~L., Franzen, T.~M.~O., Davies, R.~D., et al.\ 2009, MNRAS, 400, 984 

\bibitem[AMI Consortium: Davies et al.(2011)]{2011MNRAS.415.2708A} 
AMI Consortium, Davies, M.~L., Franzen, T.~M.~O., et al.\ 2011, MNRAS, 415, 2708 

\bibitem[AMI Consortium: Zwart et al.(2008)]{2008MNRAS.391.1545Z}
AMI Consortium, Zwart, J.~T.~L., Barker, R.~W., Biddulph, P., et al.\ 2008, MNRAS, 391, 1545 

\bibitem[Bonafede et al.(2009)]{2009A&A...503..707B}
Bonafede, A., Feretti, L., Giovannini, G., et al.\ 2009, A\&A, 503, 707 

\bibitem[Bonafede et al.(2010)]{2010A&A...513A..30B}
Bonafede, A., Feretti, L., Murgia, M., et al.\ 2010, A\&A, 513, A30 

\bibitem[{{Briggs}(1995)}]{briggs_phd}
{Briggs}, {D.~S.} 1995, PhD thesis, New Mexico Tech, USA

\bibitem[Brunetti \& Jones(2014)]{2014arXiv1401.7519B}
Brunetti, G., \& Jones, T.~W.\ 2014, arXiv:1401.7519 

\bibitem[Bykov et al.(2008)]{2008ApJ...689L.133B}
Bykov, A.~M., Uvarov, Y.~A., \& Ellison, D.~C.\ 2008, ApJL, 689, L133 

\bibitem[{Dressler}(1980)]{1980ApJ...236..351D}
{Dressler}, A.\ 1980, ApJ, 236, 351 

\bibitem[{{Drury}(1983)}]{1983RPPh...46..973D}
{Drury}, {L.~O.} 1983, Reports on Progress in Physics, 46, 973

\bibitem[{Ensslin} {et~al.}(1998)]{1998A&A...332..395E}
{Ensslin}, {T.~A.}, {Biermann}, {P.~L.}, {Klein}, {U.}, \& {Kohle}, {S.} 1998, A\&A, 332, 395

\bibitem[Feretti et al.(2001)]{2001A&A...373..106F}
Feretti, L., Fusco-Femiano, R., Giovannini, G., \& Govoni, F.\ 2001, A\&A, 373, 106 

\bibitem[{Feretti} {et~al.}(2012)]{2012A&ARv..20...54F}
{Feretti}, {L.}, {Giovannini}, {G.}, {Govoni}, {F.}, \& {Murgia}, {M.} 2012, A\&Ar, 20, 54 

\bibitem[Giacintucci et al.(2008)]{2008A&A...486..347G}
Giacintucci, S., Venturi, T., Macario, G., et al.\ 2008, A\&A, 486, 347 


\bibitem[Keshet et al.(2003)]{2003ApJ...585..128K} 
Keshet, U., Waxman, E., Loeb, A., Springel, V., \& Hernquist, L.\ 2003, ApJ, 585, 128 

\bibitem[Kocevski et al.(2007)]{2007ApJ...662..224K}
Kocevski, D.~D., Ebeling, H., Mullis, C.~R., \& Tully, R.~B.\ 2007, ApJ, 662, 224 

\bibitem[{{Noordam}(2004)}]{2004SPIE.5489..817N}
{Noordam}, {J.~E.} 2004, in SPIE Conference Series, Vol. 5489, ed. J.~M. {Oschmann}, Jr., 817--825s

\bibitem[Ogrean et al.(2013)]{2013MNRAS.429.2617O}
Ogrean, G.~A., Br{\"u}ggen, M., R{\"o}ttgering, H., et al.\ 2013, MNRAS, 429, 2617 

\bibitem[{Pacholczyk}(1970)]{pacholcyzk}
{Pacholczyk}, {A.~G}, 1970, Radio astrophysics

\bibitem[Perley \& Butler(2013)]{2013ApJS..204...19P} 
Perley, R.~A., \& Butler, B.~J.\ 2013, ApJS, 204, 19 

\bibitem[Roettiger et al.(1997)]{rottiger1997}
Roettiger, K., Loken, C., Burns, J.~O.,\ 1997, ApJS, 109, 307 

\bibitem[Roettiger et al.(1999)]{1999ApJ...518..603R}
Roettiger, K., Burns, J.~O., \& Stone, J.~M.\ 1999, ApJ, 518, 603 

\bibitem[{{Sarazin}(2002)}]{2002ASSL..272....1S}
{Sarazin}, {C.~L.} 2002, in Astrophysics and Space Science Library, Vol. 272, 1--38

\bibitem[Stroe et al.(2013)]{2013A&A...555A.110S}
Stroe, A., van Weeren, R.~J., Intema, H.~T., et al.\ 2013, A\&A, 555, A110 

\bibitem[{van~Weeren} {et~al.}(2010)]{2010Sci...330..347V}
{van~Weeren}, {R.~J.}, {R{\"o}ttgering}, {H.~J.~A.}, {Br{\"u}ggen}, {M.}, \& {Hoeft}, {M.} 2010, Science, 330, 347
\end{thebibliography}

\label{lastpage}
\nocite{*}
\end{document}